\documentclass[aps,prd,twocolumn,superscriptaddress,nofootinbib,floatfix,longbibliography]{revtex4-2}
\pdfoutput=1

\usepackage{amsmath,amsfonts,amssymb}
\usepackage{graphicx}
\usepackage{booktabs}
\usepackage{color}
\usepackage[normalem]{ulem}

\usepackage{hyperref}

\makeatletter
\let\auto@bib@innerbib\@empty
\makeatother

\begin{document}

\title{{Exploring the Supersymmetric Grand Unified Theories with Gauge Mediation at the Future Proton-Proton Colliders and Hyper-Kamiokande Experiment}\\[1mm]
}

\author{Waqas Ahmed}
\email{waqasmit@hbpu.edu.cn}
\affiliation{School of Mathematics and Physics, Hubei Polytechnic University, Huangshi 435003, China }
\author{Tianjun Li}
\email{tli@mail.itp.ac.cn}
\affiliation{School of Physics, Henan Normal University, Xinxiang 453007, P. R. China}
\author{Shabbar Raza}
\email{shabbar.raza@fuuast.edu.pk}
\affiliation{Department of Physics, Federal Urdu University of Arts, Science and Technology, Karachi 75300, Pakistan}

\begin{abstract}

As a continuation of our previous work, with the grand desert hypothesis, we prove that we can probe the Supersymmetric Grand Unified Theories (GUTs) with gauge meidation at the future proton-proton collider and Hyper-Kamiokande experiment. For the GUT scale $M_{\rm GUT}$ less than or equal to $0.84\times10^{16}$~GeV, the proton lifetime for dimension-six proton decay via heavy gauge boson exchange is less than or around $5\times10^{34}$ years, and thus we can probe it at the Hyper-Kamiokande experiment. For $ M_{\rm GUT} \ge 0.84 \times 10^{16}$ GeV, the lightest neutralino is the Next to the Lightest Supersymmetric Particle (NLSP), as well as the upper mass bounds on the gluino and first two generation squarks are 8.64 TeV and 11.82 TeV, respectively. Therefore, the gluino and first two generation squarks can be probed at the future 100 TeV proton-proton collider such as ${\rm FCC}_{\rm hh}$ and SppC. 
Moreover, we discuss the mass spectra for the colored sparticles, sleptons, pseudo-scalar Higgs and electroweakinos. If the sparticles were discovered, it might give us hint on the messenger scale, the effective supersymmetry breaking scale, and the renormalization group evolution of the supersymmetry breaking soft terms.

\end{abstract}
\maketitle

\section{Introduction}

Supersymmetry (SUSY) remains one of the most compelling extensions of the Standard Model (SM), providing an elegant solution to the gauge hierarchy problem and offering a framework in which the gauge couplings can naturally approach unification at high-energy scale ~\cite{gaugeunification}. In addition, the supersymmetric SMs (SSMs) with $Z_2$ $R$-parity provide the viable dark matter (DM) candidates, {\it i.e.},the Lightest SUSY Particles (LSPs) ~\cite{Jungman:1995df}, and establish a connection between physics at the electroweak (EW) scale. It should also be noted that the gauge coupling unification strongly hints the Grand Unified Theories (GUTs)~\cite{Georgi:1974sy,Pati:1974yy,Fritzsch:1974nn,Georgi:1974my}. Moreover, one can construct SUSY GUTS from superstring theory and thus provide possible ultraviolet completions above the grand unification scale $M_{GUT}$.

After painstaking waiting for the accumulation of data in Run-1, Run-2 and Run-3, we have no clue of any superpartner of the SM particles. Not disappointed by the situation, one can put stringent bounds on the SUSY particle spectra, and obtain that  the lower mass bounds on gluino,
the first two generation of squarks, stop, and sbottom are  2.3 TeV, 1.9 TeV, 1.25 TeV, and 1.5 TeV, respectively~\cite{ATLAS-SUSY-Search, Aad:2020sgw, Aad:2019pfy, CMS-SUSY-Search-I, CMS-SUSY-Search-II}.
Definitely, this situation also suggests the possible existence of the EW fine-tuning problem (EWFT). But one can find some promising solutions~\cite{Dimopoulos:1995mi, Cohen:1996vb, Kitano:2006gv, LeCompte:2011cn, Fan:2011yu, Kribs:2012gx,
Baer:2012mv, Ding:2015epa,Batell:2015fma,Leggett:2014hha,Du:2015una,Li:2015dil,Ellis:1986yg,Barbieri:1987fn}.

These natural SSMs generically predict some relatively light sparticles, for
instance, Bino, Higgsino, stop, gluino, and sleptons, etc. On the other hand, to explore the new physics beyond the SM, there are
a few proposals for the future proton-proton (pp) colliders, for instance,
the ${\rm FCC}_{\rm hh}$~\cite{Benedikt:2018csr}  and SppC~\cite{CEPC-SPPCStudyGroup:2015csa}.
Now the pertinent question is, can one probe the supersymmetry,
at the ${\rm FCC}_{\rm hh}$ and SppC. The simplest answer could be no since the sparticles seem to be very heavy.
The next interesting question is whether we can probe the supersymmetric GUTs at the future experiments
even if we have to face the SUSY EWFT problem. Surprisingly, the answer is yes. 
The SUSY searches at the 100~TeV pp colliders have been reported previously~\cite{Cohen:2013xda, Arkani-Hamed:2015vfh,
Fan:2017rse, Golling:2016gvc}. For the integrated luminosity 30~${\rm ab}^{-1}$,
Wino via Bino decay, gluino ${\tilde g}$ via heavy flavor decay, gluino via light flavor decay,
first-two generation squarks ${\tilde q}$, and stop can be discovered for their masses
up to about 6.5~TeV, 11~TeV, 17~TeV, 14~TeV, and 11~TeV, respectively.
Moreover, if the gluino and first-two generation squark masses are similar, they can be probed up to 20 TeV.

In previous studies~\cite{Ahmed:2020qdt,Ahmed:2022fjw}, with the grand desert hypothesis, the authors showed that
the supersymmetric GUTs can be probed at the future proton-proton (pp) colliders and
Hyper-Kamiokande experiment. For the GUTs with the GUT scale $M_{GUT} \le 1.0\times 10^{16}$~GeV,
 the dimension-six proton decay via heavy gauge boson exchange can be probed
at the Hyper-Kamiokande experiment. Moreover, for  $M_{GUT} \ge 1.0\times 10^{16}$~GeV,
it is further shown that the GUTs with anomaly and gauge mediated supersymmetry breakings are well within the reaches of
the future 100 TeV pp colliders such as the ${\rm FCC}_{\rm hh}$ and SppC,
and the supersymmetric GUTs with gravity mediated supersymmetry breaking can be probed at the future 160 TeV pp collider.

In this work, we continue our program to study the GUTs at the future 100 TeV proton-proton colliders and Hyper-Kamiokande experiment. In particularly, we explore the supersymmetric GUTs with gauge mediated supersymmetry breaking (GMSB) in the minimal realization in details~\cite{Dine:1993yw,Giudice:1998bp,Meade:2008wd}. The Hyper-Kamiokande experiment is expected to probe the proton partial lifetime in the $p\rightarrow e^{+}\pi^{0}$ channel above $10^{35}$ years~\cite{Hyper-Kamiokande:2018ofw,Hyper-KamiokandeProto-:2015xww}. For the conventional dimension-six gauge-boson-mediated proton decay, the partial lifetime scales are approximately $\tau(p\rightarrow e^{+}\pi^{0})\propto M_X^4$, and we identify $M_X\simeq M_{\rm GUT}$ \cite{Li:2010dp,Nath:2006ut}. Taking into account the theoretical uncertainty associated with the proton-decay prediction, we use the reference $M_{\rm GUT}$ value $0.84\times10^{16}$ GeV, corresponding in our reference normalization to $\tau(p\to e^+\pi^0)\simeq 5 \times10^{34}$ years.
Therefore,  for $\rm M_{\rm GUT} \le 0.84 \times 10^{16}$ GeV, the SUSY GUTs with gauge mediation can be explored at the Hyper-Kamiokande experiment.
On the other hand, for $ M_{\rm GUT} \ge 0.84 \times 10^{16}$ GeV, the lightest neutralino is the Next to the Lightest Supersymmetric Particle (NLSP), as well as the upper mass bounds on the gluino and first two generation squarks are 8.64 TeV and 11.82 TeV, respectively. Therefore, the gluino and first two generation squarks can be probed at the future 100 TeV proton-proton collider such as ${\rm FCC}_{\rm hh}$ and SppC. 
Furthermore, we discuss the mass spectra for the colored sparticles, sleptons, pseudo-scalar Higgs and electroweakinos. If the sparticles were discovered, it might give us hint on the messenger scale, the effective supersymmetry breaking scale, and the renormalization group evolution of the supersymmetry breaking soft terms.

This paper is organized as follows. In Section II, we study the supersymmetric GUTs with gauge mediated supersymmetry breaking in details. In Section III, we discuss numerical analysis and constraints. Section IV covers numerical results and spectrum analysis. Our conclusion is given in Section V.

\section{Probing the Supersymmetric Grand Unified Theories with Gauge Mediation at the Future Proton-Proton Colliders }\label{sec:scan}
The GMSB is a method of communicating SUSY breaking to the SSMs from the hidden sector through the SM gauge interactions.
We work with the minimal GMSB framework and basic parameters are \cite{Bagger:1996bt, Ambrosanio:2000cs,Giudice:1998bp}
\begin{equation}
M_{mess}, \Lambda=\frac{F}{M_{\rm mess}}
\label{eq:Lambda}
\end{equation}
as well as $\tan\beta$ and $\operatorname{sign}(\mu)$.
It should be noted that $\Lambda$ sets the characteristic scale of the soft supersymmetry-breaking parameters
transmitted to the MSSM sector.

The defining feature of gauge mediation is that gaugino and scalar masses are
generated radiatively through messenger loops. At the messenger scale, the
gaugino masses appear at one-loop order and are given by

\begin{equation}
M_a(M_{\rm mess})
=
N_{\rm mess}
\frac{\alpha_a(M_{\rm mess})}{4\pi}
\Lambda\,g(x),
\label{eq:gaugino}
\end{equation}

where $a=1,2,3$ corresponds to the three SM gauge groups,
$N_{\rm mess}$ is the messenger index, and $g(x)$ is the messenger threshold
function with $x=F/M_{\rm mess}^2$. In the limit of small messenger splitting,
$x\ll1$, this function approaches unity and the leading behavior becomes

\begin{equation}
M_a\propto\alpha_a .
\end{equation}

This relation represents one of the characteristic predictions of gauge
mediation: the relative hierarchy among gaugino masses is determined by the
SM gauge couplings rather than by arbitrary high-scale boundary
conditions.

The scalar sector receives supersymmetry-breaking masses at the two-loop
level. For a scalar field $\tilde f$, the soft mass at the messenger scale is

\begin{equation}
m_{\tilde f}^{2}(M_{\rm mess})
=
2N_{\rm mess}\Lambda^2
\sum_a
C_a^{\tilde f}
\left(
\frac{\alpha_a(M_{\rm mess})}{4\pi}
\right)^2
f(x),
\label{eq:scalar}
\end{equation}

where $C_a^{\tilde f}$ denotes the quadratic Casimir invariant of the scalar
representation under the corresponding gauge group. The function $f(x)$
contains threshold corrections and approaches unity in the limit of small
messenger splitting. Since these masses depend only on gauge quantum numbers,
gauge mediation naturally suppresses flavour violation.

The soft parameters generated at the messenger scale are evolved to the
electroweak scale through renormalization group equations. The running of the
gauge couplings is described by

\begin{equation}
\frac{d g_a}{d\ln Q}
=
\frac{b_a}{16\pi^2}g_a^3 ,
\label{eq:RGE}
\end{equation}

while the gaugino masses satisfy

\begin{equation}
\frac{dM_a}{d\ln Q}
=
\frac{2b_a}{16\pi^2}g_a^2M_a .
\label{eq:RGEgaugino}
\end{equation}

The renormalization group evolution is crucial because the low-energy spectrum
is not determined only by the messenger-scale relations. In particular, the
relative running of bino, wino, and slepton masses controls the identity of the
next-to-lightest supersymmetric particle (NLSP), which is the central
phenomenological feature of gauge-mediated models.

A distinctive prediction of GMSB is that the gravitino is typically the
lightest supersymmetric particle (LSP). The gravitino mass is related to the
fundamental supersymmetry-breaking scale through

\begin{equation}
m_{3/2}
=
\frac{F_{\rm tot}}
{\sqrt{3}M_{\rm Pl}},
\label{eq:gravitino}
\end{equation}

where $F_{\rm tot}$ represents the total supersymmetry-breaking order
parameter. Because the gravitino interaction strength is suppressed by the
Planck scale, it naturally remains much lighter than the MSSM superpartners.

The phenomenology of gauge mediation is therefore governed by the properties
of the NLSP. In the parameter region studied in this work, the NLSP is
typically either the lightest neutralino or the lighter stau. The neutralino
NLSP scenario is characterized by the decay

\begin{equation}
\tilde{\chi}^{0}_{1}
\rightarrow
\gamma/Z/h+\tilde G ,
\label{eq:neutralino_decay}
\end{equation}

whereas a stau NLSP follows

\begin{equation}
\tilde{\tau}_{1}
\rightarrow
\tau+\tilde G .
\label{eq:stau_decay}
\end{equation}

The transition between these two possibilities is determined by the mass
ordering

\begin{equation}
m_{\tilde{\tau}_1}
=
m_{\tilde{\chi}^{0}_{1}} .
\label{eq:NLSP_boundary}
\end{equation}

The position of this boundary depends on the messenger scale, the effective
supersymmetry-breaking scale, and the renormalization group evolution of the
soft parameters. Consequently, the neutralino--stau NLSP plane provides a
direct probe of the underlying structure of the gauge-mediated model.

\section{Numerical Analysis and Constraints}
\label{sec:numerical}

To investigate the phenomenological consequences by imposing the $\rm M_{GUT}$ condition and other experimental constraints on the gauge mediated supersymmetry breaking parameter space, we perform a comprehensive numerical analysis in the minimal GMSB framework. The model implementation and spectrum calculation are performed using the package \textsc{SARAH}4.14.3~\cite{Porod:2003um}, which automatically generates the
corresponding \textsc{SPheno}4.0.4~\cite{Staub:2008uz} module for calculating the supersymmetric
particle spectrum, decay properties, and low-energy observables. The generated
\textsc{SPheno} code is then interfaced with our scanning framework to perform
a random exploration of the GMSB parameter space.

In the minimal realization of gauge mediation, as described before, the relevant parameters
controlling the low-energy supersymmetric spectrum are the effective
supersymmetry-breaking scale $\Lambda$, the messenger scale $M_{\rm mess}$,
the messenger multiplicity $N_{\rm mess}$, the ratio of Higgs vacuum
expectation values $\tan\beta$, and the sign of the Higgsino mass parameter
$\mu$. The messenger fields are assumed to form complete multiplets of the
grand unified group $SU(5)$ in order to preserve gauge coupling unification.
For simplicity, we consider one pair of messenger multiplets,

\begin{equation}
\Phi+\bar{\Phi} = {\bf 5}+\overline{\bf 5},
\end{equation}

corresponding to

\begin{equation}
N_{\rm mess}=1 .
\end{equation}

The soft supersymmetry-breaking parameters are generated at the messenger
scale and subsequently evolved to the electroweak scale through the full
renormalization group equations (RGEs) implemented in \textsc{SPheno}. The scan is
performed over the following minimal GMSB parameter space:

\begin{equation}
5\times10^{5}\ {\rm GeV}
\leq
\Lambda
\leq
10^{7}\ {\rm GeV},
\end{equation}

\begin{equation}
2\Lambda
\leq
M_{\rm mess}
\leq
10^{15}\ {\rm GeV},
\end{equation}

and

\begin{equation}
2\leq\tan\beta\leq60 ,
\end{equation}

with

\begin{equation}
\mu>0 .
\end{equation}

The top-quark mass is fixed to~ \cite{:2009ec}

\begin{equation}
m_t=173.2~{\rm GeV}.
\end{equation}
We would like to draw attention of the reader to the fact that the requirement of radiative electroweak symmetry breaking (REWSB)~\cite{Ibanez:1982fr} puts an important theoretical constraint on the parameter space.

The data points collected all satisfy the requirement of REWSB. In addition, after collecting the data, we impose the SM-like Higss mass bound $m_{h}=[123-127]$ GeV, mass bounds on all the sparticles \cite{Agashe:2014kda}, and the constraints from rare decay processes $B_{s}\rightarrow \mu^{+}\mu^{-} $ \cite{Aaij:2012nna}, $b\rightarrow s \gamma$ \cite{Amhis:2012bh}, and $B_{u}\rightarrow \tau\nu_{\tau}$ \cite{Asner:2010qj}.
More explicitly, we set
\begin{align}
m_{\tilde g} &\geq 2.2~{\rm TeV}, \\
0.8\times10^{-9}
&\leq {\rm BR}(B_s\to\mu^+\mu^-)
\leq 6.2\times10^{-9}\quad(2\sigma), \\
2.99\times10^{-4}
&\leq {\rm BR}(b\to s\gamma)
\leq 3.87\times10^{-4}\quad(2\sigma), \\
0.15
&\leq
\frac{{\rm BR}(B_u\to\tau\nu_\tau)_{\rm MSSM}}
     {{\rm BR}(B_u\to\tau\nu_\tau)_{\rm SM}}
\leq 2.41\quad(3\sigma).
\end{align}

\subsection{Determination of GUT scale}
\label{sec:gutscale}
Since the calculation of the grand unification scale is not directly provided
by all spectrum generators, we determine the effective unification scale by
tracking the running of the gauge couplings. We define

\begin{equation}
\alpha^{-1}_{12}(Q)
=
\alpha_1^{-1}(Q)
-
\alpha_2^{-1}(Q),
\end{equation}

where the gauge couplings are obtained from the \textsc{SPheno} output \cite{Porod:2003um}. The evolution of the gauge couplings is governed by

\begin{equation}
\frac{d\alpha_i^{-1}}{d\ln Q}
=
-\frac{b_i}{2\pi},
\end{equation}

at one-loop order \cite{Martin:1993zk}. A reference unification curve is obtained from a benchmark
CMSSM point with

\begin{equation}
M_{\rm GUT}\simeq10^{16}\ {\rm GeV}.
\end{equation}

The reference function

\begin{equation}
f(Q)
\equiv
\alpha^{-1}_{12}(Q)
\end{equation}

is fitted numerically. For each GMSB point, the values of
$\alpha^{-1}_{1}(M_{\rm mess})$ and
$\alpha^{-1}_{2}(M_{\rm mess})$ are extracted from \textsc{SPheno} and
compared with the reference running. Points satisfying

\begin{equation}
\alpha^{-1}_{12}(M_{\rm mess})>f(M_{\rm mess})
\end{equation}

correspond to

\begin{equation}
M_{\rm GUT}>10^{16}\ {\rm GeV},
\end{equation}

whereas

\begin{equation}
\alpha^{-1}_{12}(M_{\rm mess})<f(M_{\rm mess})
\end{equation}

indicates

\begin{equation}
M_{\rm GUT}<10^{16}\ {\rm GeV}.
\end{equation}

This procedure allows us to correlate the low-energy supersymmetric spectrum
with the underlying high-scale unification properties of the gauge-mediated
scenario.
\subsection{Constraint from Proton Decay Experiment}
For dimension-six proton decay induced by heavy $X/Y$ gauge-boson exchanges, the
partial lifetime has the characteristic fourth-power dependence \cite{Hisano:2000dg,Dutta:2016jqn}
\begin{align}
\tau(p\to e^+\pi^0)
\simeq{}&10^{35}\ {\rm yr}
\left(\frac{M_{\rm X/Y}}{10^{16}\ {\rm GeV}}\right)^4
\nonumber\\
&\times
\left(\frac{0.04}{\alpha_{\rm GUT}}\right)^2
\left(\frac{2.5}{A_R}\right)^2 .
\label{eq:proton-lifetime-benchmark}
\end{align}
where the numerical normalization represents a standard reference choice for
the chiral and lattice-QCD matrix elements, and $A_R$ denotes the
operator-renormalization factor
\cite{Hisano:2022proton,Yoo:2021qcd}.  Solving Eq.~\eqref{eq:proton-lifetime-benchmark}
for the heavy-gauge-boson mass gives, for the reference values,
\begin{equation}
\frac{M_{\rm X/Y}}{10^{16}\ {\rm GeV}}
\simeq
\left(\frac{\tau}{10^{35}\ {\rm yr}}\right)^{1/4}.
\label{eq:MX-from-tau}
\end{equation}
Consequently,
\begin{align}
\tau=5\times10^{34}\ {\rm yr}
&\quad\Longleftrightarrow\quad
M_X\simeq0.84\times10^{16}\ {\rm GeV}.
\label{eq:proton-benchmark-map}
\end{align}
This is precisely the  red color used in the following figures.
For comparison, the Super--Kamiokande limit on dimension-six proton decay
$\tau(p\to e^+\pi^0)>2.4\times10^{34}$~yr maps to
$M_{\rm X/Y}\gtrsim0.70\times10^{16}$~GeV under the same normalization
\cite{Takenaka:2020proton}. Also, a sensitivity of $10^{35}$~yr corresponds to
$M_{\rm X/Y}\sim 1.0\times 10^{16}$~GeV and motivates the Hyper--Kamiokande target region
\cite{HyperK:2018design}.

At the future Hyper-Kamiokande experiment, we can probe the proton lifetime
at least above $1.0 \times 10^{35}$ years~\cite{HyperK:2018design}.
Thus, for $M_{\rm X/Y} \le 0.84\times 10^{16}~{\rm GeV}$,  the GUTs with gauge mediation is within the reach of 
the future Hyper-Kamiokande experiment. 
For more detail about GUTs and proton decays, 
 see~\cite{Hisano:2013cqa, Ellis:2015jwa, Pokorski:2017ueo,  Pokorski:2019ete,Tobe:2003yj,Goto:1998qg, Murayama:2001ur,Liu:2013ula,Hisano:2013exa,Babu:2020ncc}.
Moreover, to clarify the subtle point, we want to emphasize that the mass of the heavy gauge bosons $X_{\mu}/Y_{\mu}$
is smaller than or equal to the GUT scale  $M_{GUT}$. Thus, for $M_{\rm GUT} \le 0.84\times 10^{16}~{\rm GeV}$,  the GUTs with gauge mediation can be probed at the future Hyper-Kamiokande experiment.


\section{Numerical Results and Spectrum Analysis}
\label{sec:numerical_results}
After collecting data, we impose the constraints of Sec.~\ref{sec:numerical}.  We organize the
sample in a common hierarchy used in all figures:

The figure legends actually use mutually exclusive $\rm M_{\rm GUT}$ bins after the basic phenomenological constraints:
\begin{itemize}
\item gray: generated points before phenomenological cuts (apart from the technical requirements needed to obtain a consistent spectrum);
\item yellow: points satisfying the basic Higgs-mass, sparticle-mass, and $B$-physics constraints with $M_{\rm GUT}<0.84\times10^{16}$~GeV;
\item red: points satisfying the basic constraint with $\rm M_{\rm GUT}>0.84\times10^{16}$~GeV.
\end{itemize}

Prospective studies of $100$~TeV proton-proton colliders find that the gluino reach can vary widely with the decay topology and spectrum.  The representative
 $11$, and $17$~TeV horizontal lines shown below are therefore not two
simultaneous exclusion bounds.  They are visual guides associated with
different search assumptions, such as heavy- and
light-flavor gluino decays
\cite{Cohen:2013xda,Arkani-Hamed:2015vfh,Fan:2017rse,Golling:2016gvc}.

\subsection{Messenger-scale parameter space}

\begin{figure*}[t]
\centering
\includegraphics[width=0.85\textwidth]{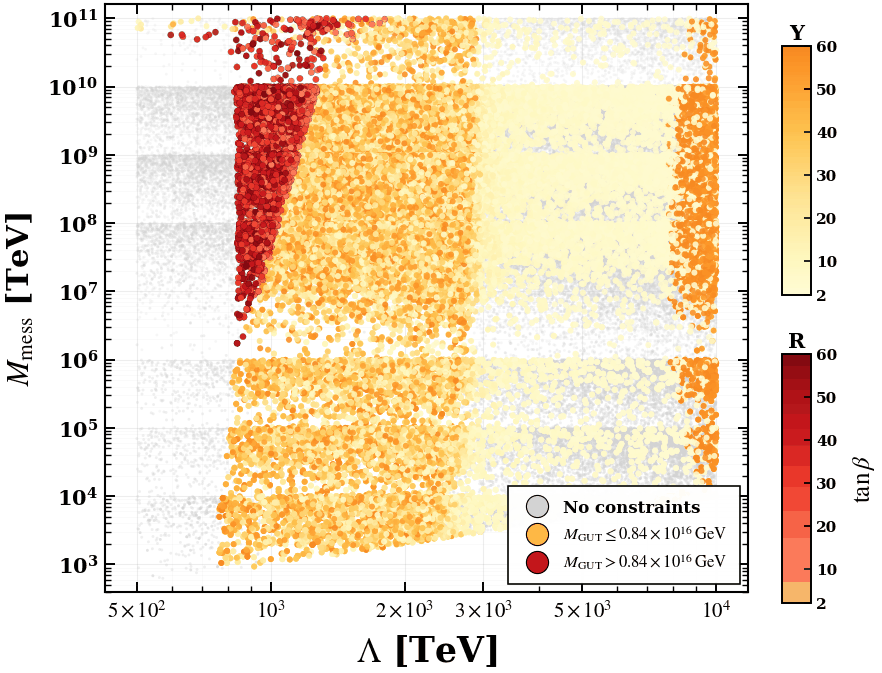}
\caption{
Allowed parameter space in the $\Lambda$-$M_{\rm mess}$ plane.  Gray points
show the raw scan before phenomenological cuts.  The colored points satisfy
the basic Higgs, sparticle-mass, and flavor constraints and are separated into
two mutually exclusive GUT-scale bins: yellow for
$\rm M_{\rm GUT}<0.84\times10^{16}$~GeV, and red for
$\rm M_{\rm GUT}>0.84\times10^{16}$~GeV.  Within each colored bin, the corresponding
color gradient gives $\tan\beta$.}
\label{fig:messenger_plane}
\end{figure*}

Figure~\ref{fig:messenger_plane} displays the fundamental GMSB plane.  The
effective scale $\Lambda$ fixes the leading size of the soft terms,
\begin{equation}
M_a\propto\Lambda,
\qquad
m_{\widetilde f}^{2}\propto\Lambda^2,
\label{eq:soft-scaling-results}
\end{equation}
whereas $M_{\rm mess}$ controls the length of the renormalization-group
interval between mediation and the electroweak scale.  

The most important feature of Fig.~\ref{fig:messenger_plane} is the progressive
horizontal bands of the allowed region.  The yellow $\rm M_{\rm GUT}<0.84\times10^{16}$~GeV bin extends
to large $\Lambda \sim 8\times 10^{2}-10^{4}$ TeV, while the  red selections favor
successively smaller values of $\Lambda$. For instance, the red band confined in the range $8\times10^{2}$ to $1\times10^{3}$ TeV.  Since both gaugino and scalar masses
increase with $\Lambda$, this immediately explains why the higher-$\rm M_{\rm GUT}$
bins populate the lower-mass regions of all subsequent
spectrum plots.  This trend is a numerical property of the threshold pattern
in the present minimal-GMSB scan; it should not be promoted to a theorem for
all gauge-mediated or all GUT models.

By contrast, each colored class spans a broad interval of $\rm M_{\rm mess}$ except the red points where solutions are confined to $10^{6}$ to $10^{11}$ TeV.  On the other hand yellow band seems to be independent of $\rm M_{mess}$. The messenger
fields form a complete ${\bf5}+\overline{\bf5}$ multiplet of $SU(5)$, whose
one-loop contribution shifts the three GUT-normalized beta-function
coefficients equally and therefore preserves the one-loop differences that
control the gauge-coupling crossing.  The messenger scale still affects the
low-energy spectrum through the length of RGEs evolution, two-loop effects, and
threshold corrections, but its impact on $\rm M_{\rm GUT}$ is less direct than the
impact of the superpartner threshold pattern controlled by $\Lambda$.

The $\tan\beta$ gradients do not reveal a unique narrow preferred value.
Nevertheless, large $\tan\beta$ enhances the bottom and tau Yukawa couplings,
which modifies the third-generation and Higgs soft parameters and can lower
the lighter-stau mass.  The color gradients should therefore be interpreted as
a secondary structure superimposed on the dominant $\Lambda$ dependence, not
as evidence that the proton-decay threshold is determined by $\tan\beta$ alone.

\subsection{Colored spectrum and Higgs-sector correlations}

\begin{figure*}[t]
\centering
\includegraphics[width=0.48\textwidth]{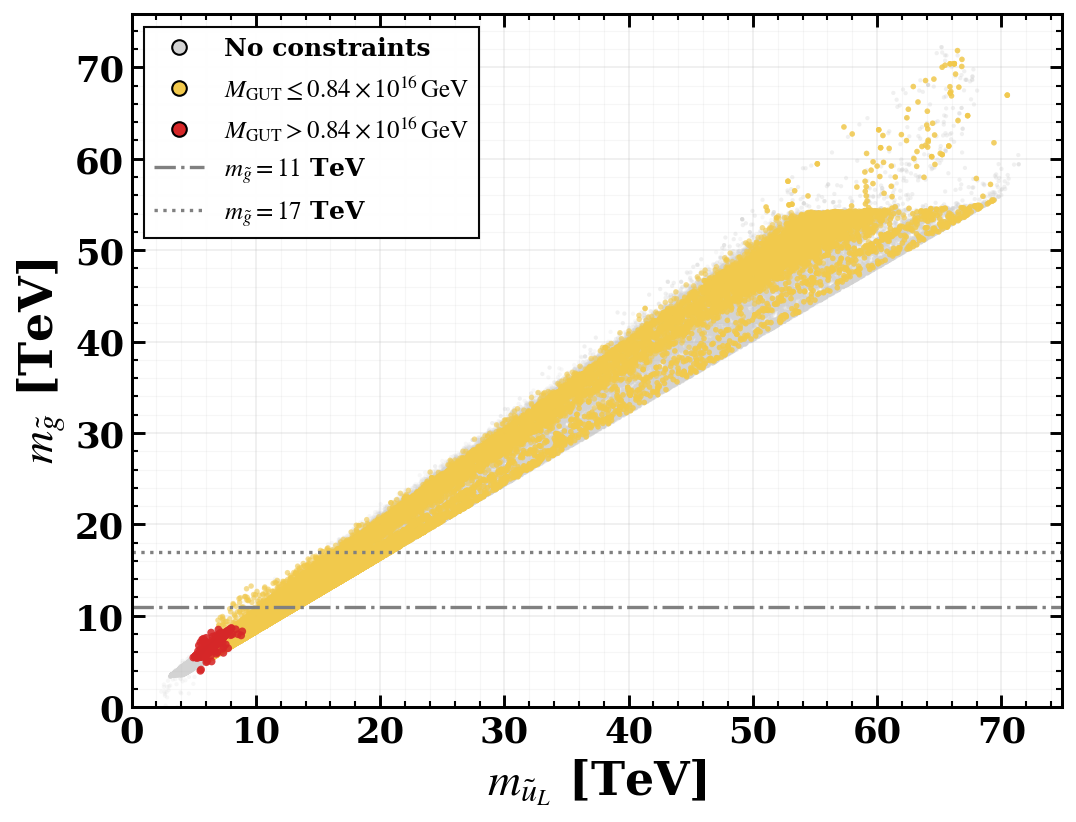}
\includegraphics[width=0.48\textwidth]{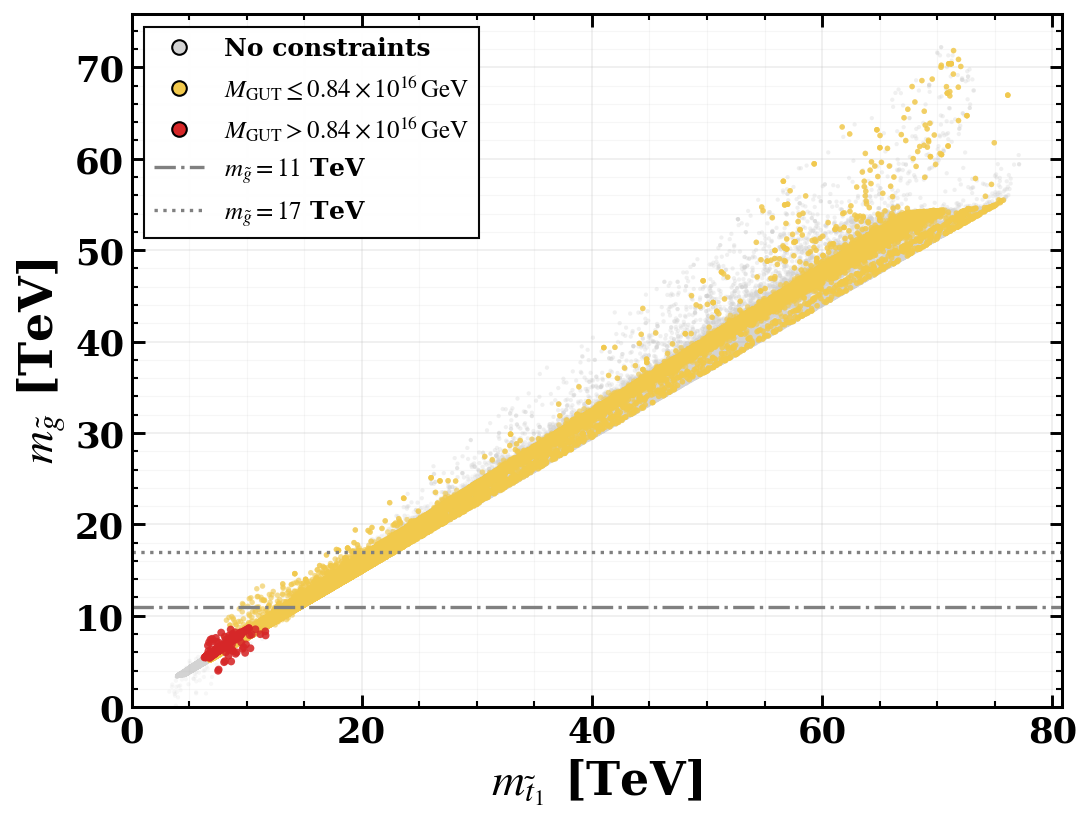}
\includegraphics[width=0.48\textwidth]{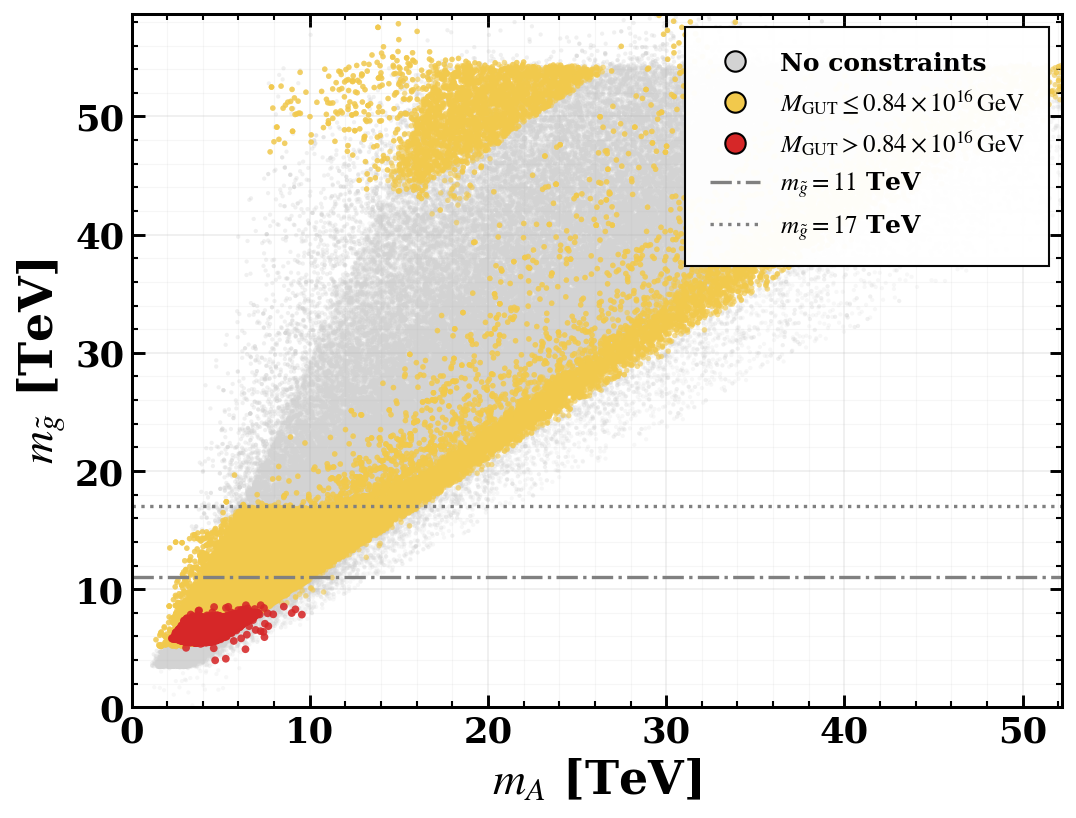}
\caption{Correlations of the gluino mass with the first/second-generation
squark mass, the lighter-stop mass, and the pseudoscalar
Higgs mass.  The point colors have the same meaning as in
Fig.~\ref{fig:messenger_plane}: gray denotes the raw scan, while the colored
points satisfy the basic constraints and are divided into the mutually exclusive
bins yellow ($\rm M_{\rm GUT}<0.84\times10^{16}$~GeV),  and red
($\rm M_{\rm GUT}>0.84\times10^{16}$~GeV).  The horizontal $11$, and $17$~TeV
lines are representative topology-dependent gluino sensitivities.}
\label{fig:spectrum}
\end{figure*}

The near-linear first/second-generation squark--gluino mass correlations in
Fig.~\ref{fig:spectrum} follows directly from gauge mediation.  Both masses are
dominated by the strong interaction,
\begin{equation}
m_{\widetilde q}^{2},\ M_3^2
\propto\alpha_3^2\Lambda^2,
\end{equation}
and the negligible light-generation Yukawa couplings preserve approximate
squark degeneracy.  The red sample is concentrated at the lowest
gluino and scalar masses, 
while the yellow sample extends further into the tens of TeV spectra. 
We find that for $\rm M_{\rm GUT} > 0.84 \times 10^{16}$ GeV (red points), the upper mass bounds on the gluino, left-handed up squark, 
right-handed up squark, left-handed down squark, and right-handed down squark are 8.64 TeV, 8.91 TeV, 11.9 TeV, 11.3 TeV, and 11.9 TeV, 
respectively. Thus, the gluino and first two-generation squarks are completely accessible at the future 100 TeV proton-proton colliders 
such as $LHC_{\rm hh}$ and SppC.
On the other hand, the orange points with $\rm M_{\rm GUT} < 0.84 \times 10^{16}$ can be probed at the Hyper-Kamiokande experiment. Moreover, the 11 TeV and 17 TeV lines for gluino masses also put bounds on the squark masses. For instance, the gluino searches through heavy flavor decays and through light flavor decays put upper bounds on the squark masses of about 14 TeV and 20 TeV, respectively. 

In the right panel, we see that the lighter stop has a broader correlation because its
mass also receives top-Yukawa and left-right mixing effects. It can also be seen that for the red points, the upper bound on the lighter stop mass is about 11.66 TeV. The orange points with $\rm M_{\rm GUT} < 0.84 \times 10^{16}$ GeV will be accessible at the Hyper-kamiokande experiment, and the gluino searches through heavy and light flavor decays put upper bounds on the lighter stop mass of about 18 TeV and 25 TeV, respectively.

The pseudoscalar Higgs mass is correlated with the overall SUSY scale through
the electroweak-symmetry-breaking conditions,
\begin{equation}
m_A^2\simeq m_{H_u}^2+m_{H_d}^2+2|\mu|^2,
\end{equation}
at tree level, with substantial loop corrections included in the numerical
spectrum. The red $m_A$ distribution should therefore not be interpreted as
an independent hard prediction; rather, it reflects the same parameter
selection that leads to the lighter colored spectrum. In the
gluino--pseudoscalar Higgs mass plane, the future 100~TeV colliders via gluino decays through heavy flavors, can probe 
pseudoscalar Higgs masses up to approximately 10~TeV in the red region. It can be seen that gluino through heavy flavor decays can cover regions up to $m_{A}\sim$ 12 TeV and via light flavor $m_{A}\sim$ 16 TeV can be probed. It should be noted that solutions with $\rm M_{\rm GUT} < 0.84 \times 10^{16}$ can also be probed at Hyper-Kamiokande experiment.

\subsection{Selepton spectrum correlations}
\begin{figure*}[t]
\centering
\includegraphics[width=0.48\textwidth]{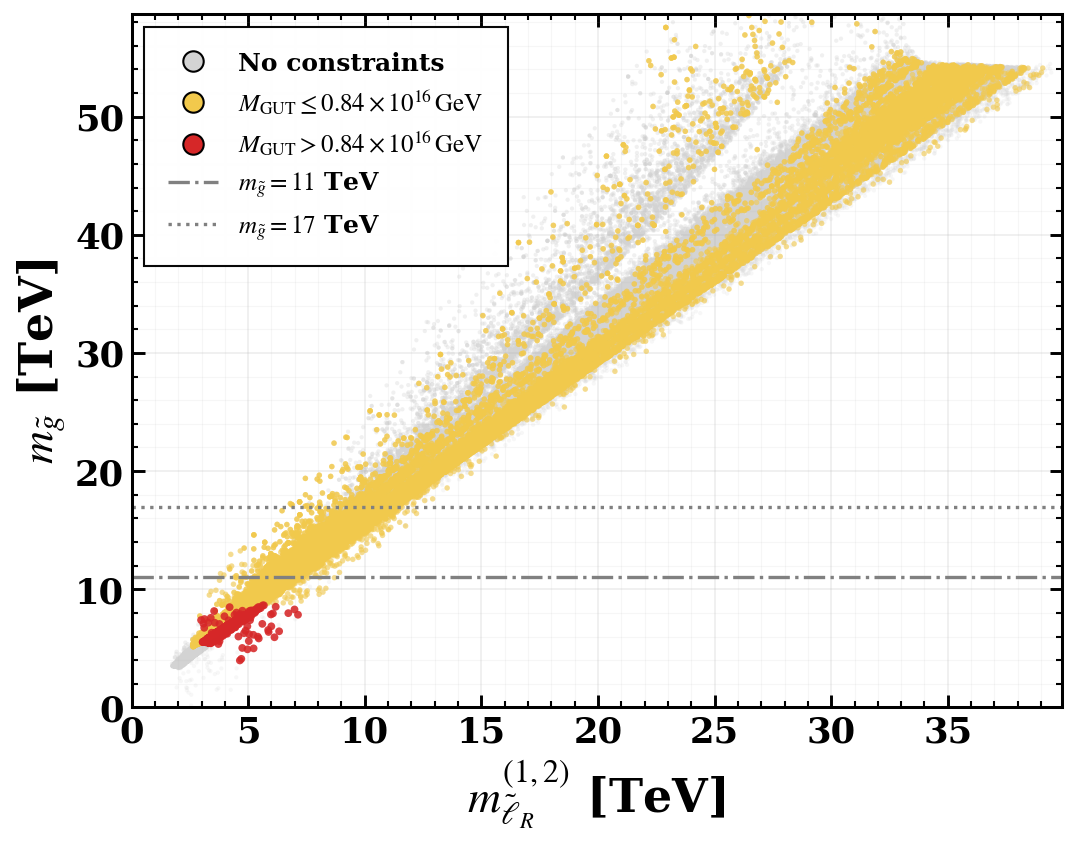}
\includegraphics[width=0.48\textwidth]{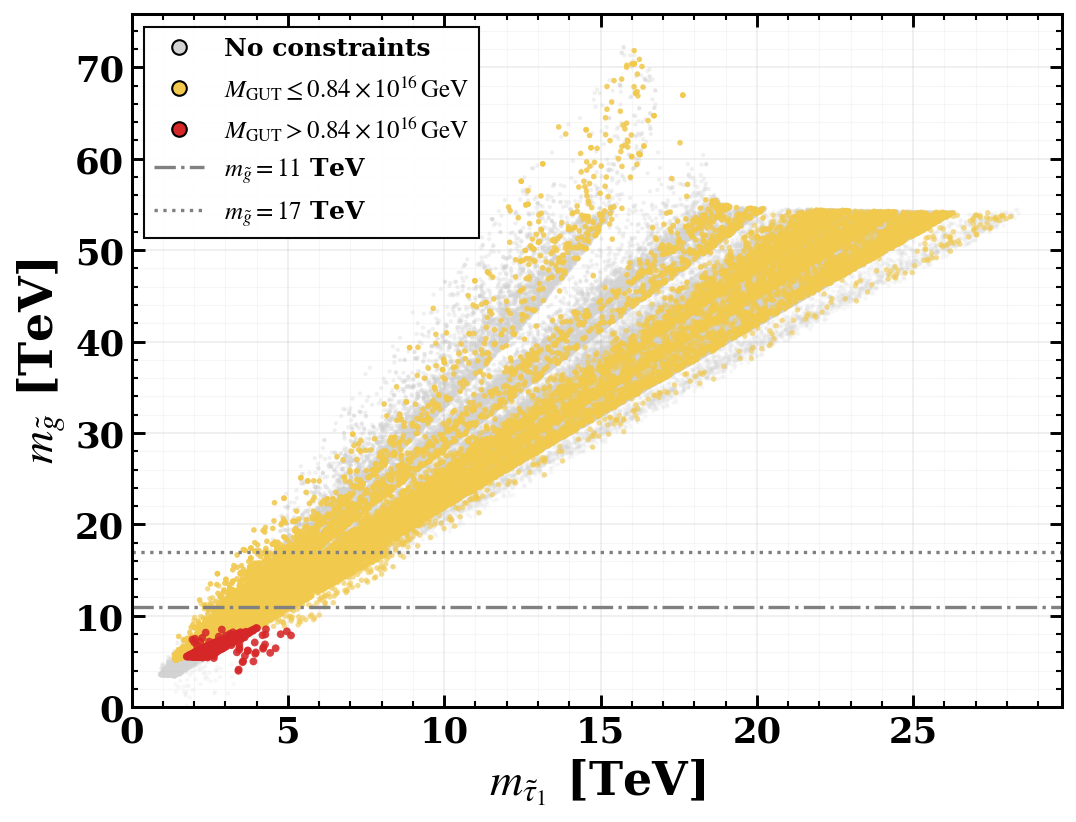}
\caption{Correlations of the gluino mass with the right-handed selectron mass, and the lighter-stau mass.  The point colors have the same meaning as in
Fig.~\ref{fig:messenger_plane}: gray denotes the raw scan, while the colored
points satisfy the basic constraints and are divided into the mutually exclusive
bins yellow ($\rm M_{\rm GUT}<0.84\times10^{16}$~GeV), and red
($M_{\rm GUT}>0.84\times10^{16}$~GeV).  The horizontal $11$, and $17$~TeV
lines are representative topology-dependent gluino sensitivities.}
\label{fig:slectron}
\end{figure*}

Figure~\ref{fig:slectron} shows that the hierarchy induced by the GUT-scale cuts is not limited to colored particles.  Because all minimal-GMSB soft terms are controlled by the same $\Lambda$, the red $\rm M_{\rm GUT}$ bin also selects the lower ends of the selepton distributions. Here we can see that upper bound on right-handed selectron mass for the red points is about 7 TeV. All the red points are accessible at the future 100 TeV colliders and the Hyper-Kamiokande experiment will probe points with $\rm M_{\rm GUT}< 0.84 \times 10^{16}$ GeV. The 11 TeV and 17 TeV lines set the upper limits on the first-two generation selectron mass of about 8 TeV and 12 TeV respectively.

On the right hand side, the stau panel has a visibly larger spread than the first-generation squark panel.  This follows from the tau Yukawa contribution and left-right mixing, which become important at large $\tan\beta$ and can reduce $m_{\widetilde\tau_1}$. In this panel we see that upper bounds on right-handed selectron for red points is about 5 TeV. The solutions with $\rm M_{\rm GUT} < 0.84 \times 10^{16}$ GeV can be probed at the at Hyper-Kamiokande experiment. The gluino searches via heavy and light flavors set the upper bounds on the lighter stau mass as 6 TeV and 9 TeV respectively.

\subsection{Electroweakinos and the identity of the NLSP}

\begin{figure*}[t]
\centering
\includegraphics[width=0.48\textwidth]{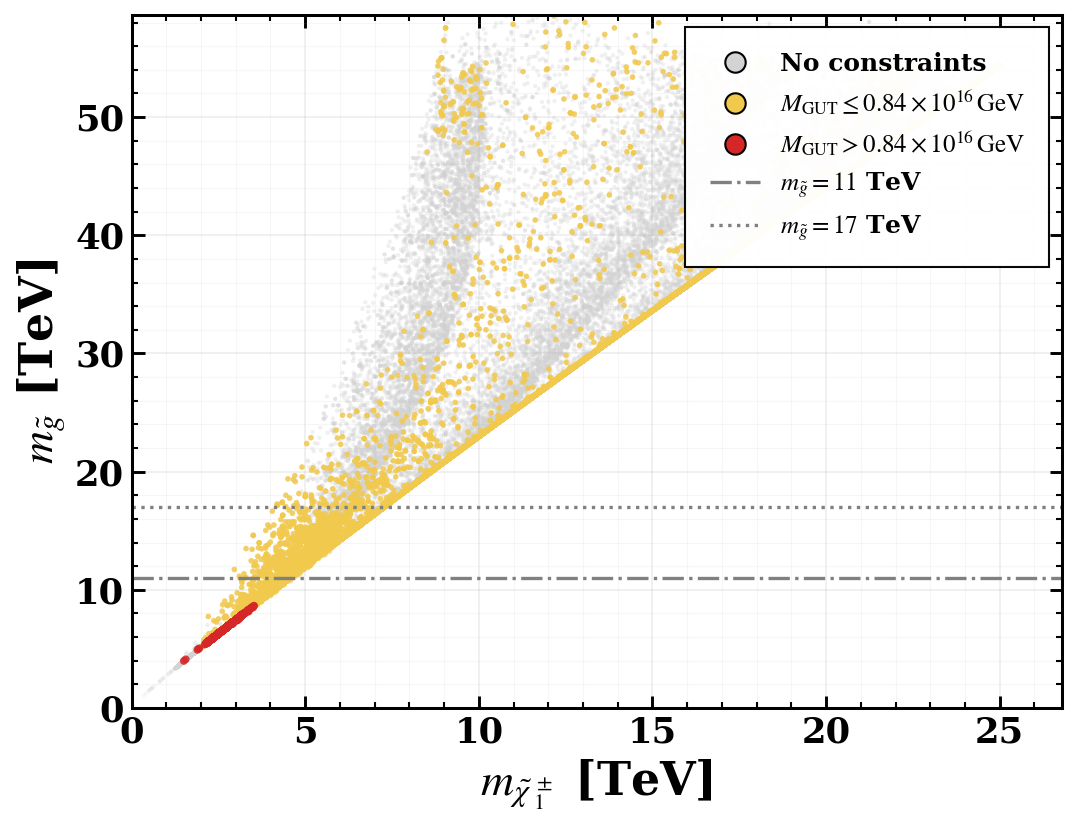}
\includegraphics[width=0.48\textwidth]{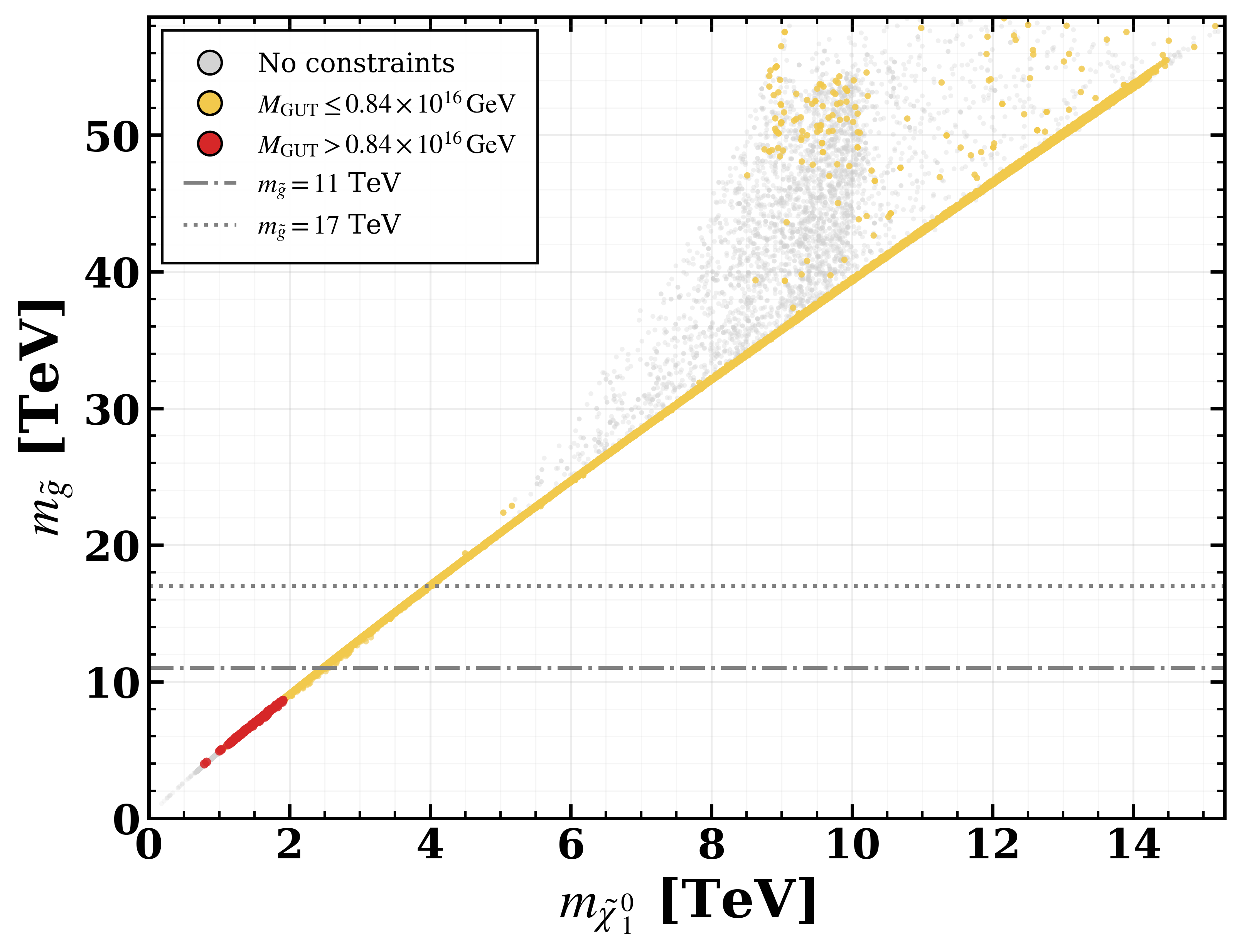}
\hfill
\includegraphics[width=0.48\textwidth]{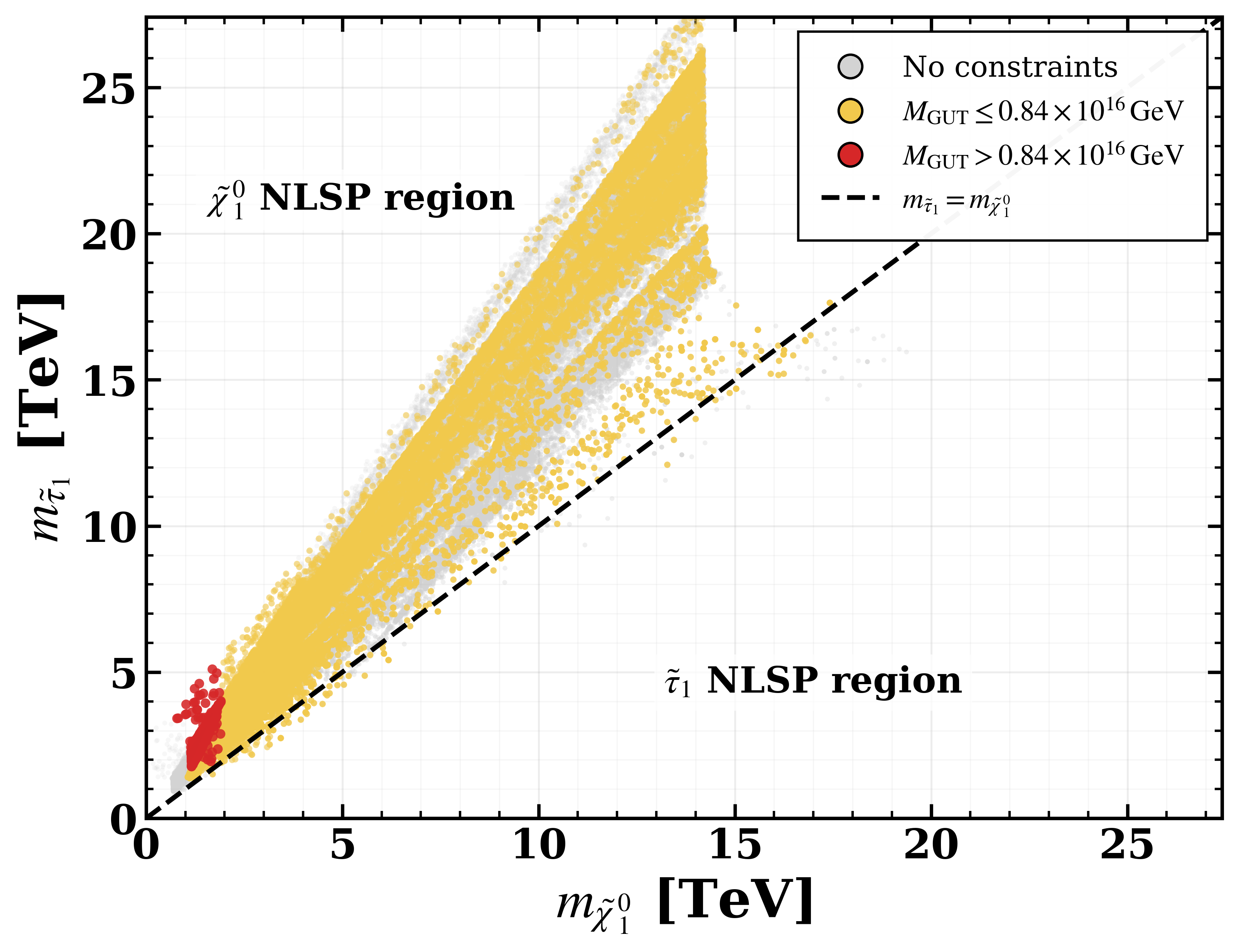}
\caption{Left: correlation between the lightest-chargino and gluino masses.  Right: correlation between the lightest-neutralino and gluino masses. Below:
the $m_{\widetilde\chi_1^0}$-$m_{\widetilde\tau_1}$ plane.  The dashed
diagonal satisfies $m_{\widetilde\tau_1}=m_{\widetilde\chi_1^0}$; points above
it have a neutralino NLSP, while points below it have a stau NLSP.  The gray,
yellow and red samples follow the same raw-scan and mutually exclusive
$M_{\rm GUT}$-bin definitions as in the preceding figures.}
\label{fig:nlsp}
\end{figure*}

The top two panels of Figure~\ref{fig:nlsp} show the correlations of the gluino mass with the chargino and neutralino masses.  In the chargino--gluino plane, the red bins extends to chargino masses of about 2.5 TeV; in the neutralino--gluino plane, the corresponding upper values are about 2~TeV.  It is evident that charginos and neutralinos of the masses mentioned above are completely accessible at the future 100 TeV proton-proton colliders such as $\rm LHC_{\rm hh}$ and SppC via gluino decays through heavy flavors. The parameter space with with $\rm M_{\rm GUT}< 0.84 \times 10^{16}$ GeV (orange solutions) is accessible at Hyper-Kamiokande experiment.
Moreover, the 11 TeV and 17 TeV lines for gluino masses put the upper mass bounds on $\chi_1^0$ and $\chi_1^{\pm}$ as
 2.5 and 5 TeV,  as well as 5 and 7 TeV, respectively. 

In order to study the NLSP in our case we need to consider some important points. The bino-like neutralino mass is generated at one loop,
\begin{equation}
M_1\simeq N_{\rm mess}\frac{\alpha_1}{4\pi}\Lambda,
\end{equation}
whereas the right-handed stau receives a two-loop gauge contribution and
important Yukawa and mixing corrections,
\begin{equation}
m_{\widetilde\tau_R}^2(M_{\rm mess})
\simeq2N_{\rm mess}\Lambda^2 C_1
\left(\frac{\alpha_1}{4\pi}\right)^2.
\end{equation}
Their different running determines which state is the NLSP. Consider the plot in the bottom panel of Fig.~\ref{fig:nlsp}, with
$m_{\widetilde\chi_1^0}$ on the horizontal axis and
$m_{\widetilde\tau_1}$ on the vertical axis, points above the diagonal obey
$m_{\widetilde\chi_1^0}<m_{\widetilde\tau_1}$ and therefore have a neutralino
NLSP.  Points below the line obey
$m_{\widetilde\tau_1}<m_{\widetilde\chi_1^0}$ and have a stau NLSP.

It can be seen that in our scans all the red points in the plot are above the diagonal line.
Thus the points with $M_{\rm GUT} > 0.84 \times 10^{16}$ (red points) favors a
neutralino NLSP in the present scan.  Moreover, it can also be noticed that almost all the blue points are above the digonal line, thus favoring NLSP neutralino. The corresponding leading signatures are
\begin{equation}
\widetilde\chi_1^0\to\gamma\widetilde G,\qquad
\widetilde\chi_1^0\to Z\widetilde G,\qquad
\widetilde\chi_1^0\to h\widetilde G,
\end{equation}
with branching fractions set by the neutralino composition and phase space.
Some yellow solutions also favor lighter stau NLSP. A lighter stau NLSP decays through
$\widetilde\tau_1\to\tau\widetilde G$ and can yield prompt taus, displaced
vertices, or heavy charged tracks depending on the fundamental SUSY-breaking
scale.  The NLSP plane therefore connects the high-scale color classification
to experimentally distinguishable GMSB signatures.

\subsection{Collider signatures associated with the two NLSP branches}
\label{sec:collider-signatures}

For a neutralino NLSP, strongly produced gluinos and squarks cascade to
$\widetilde\chi_1^0$, after which the final transition to the gravitino gives
\begin{align}
pp&\to\widetilde g\widetilde g,\,
\widetilde q\widetilde q,\,\widetilde q\widetilde g
\nonumber\\
&\to {\rm jets}+\widetilde\chi_1^0\widetilde\chi_1^0
\nonumber\\
&\to {\rm jets}+
\{\gamma\gamma,\gamma Z,ZZ,Zh,hh\}+E_T^{\rm miss}.
\label{eq:neutralino-cascade}
\end{align}
The relative rates depend on the bino, wino, and Higgsino components.  A
predominantly bino NLSP usually enhances the photonic channel, while sizable
Higgsino components can increase $Z\widetilde G$ and $h\widetilde G$ modes.
For long lifetimes, the same decay produces delayed or nonpointing photons
rather than a prompt diphoton signature
\cite{Meade:2009qv,Kawagoe:2003jv}.

\begin{table*}[t]
\centering
\scalebox{1.0}{
\begin{tabular}{lcc}
\hline
\hline
                 & Point 1 (Red) & Point 2 (Yellow) \\

\hline
$\Lambda$          & 1754331.0 & 1782445.0    \\
$M_{mess}$          & $8.471\times 10^{13}$ & $1.783\times 10^{6}$\\
$\tan\beta$        & 11.78   & 52.97\\
\hline
$M_{GUT}$         & $0.853\times 10^{16}$  & $0.592\times 10^{16}$ \\
\hline
$m_h$            & 125  & 125.6   \\
$m_H$            & 9194 & 3539   \\
$m_{A}$           & 9194 & 3539\\
$m_{H^{\pm}}$     & 9185 & 3532 \\

\hline
$m_{\tilde{\chi}^0_{1,2}}$
                 & 1801, 3342 & 3432, 3505 \\

$m_{\tilde{\chi}^0_{3,4}}$
                 & 6173, 6174 & 3518, 6303 \\

$m_{\tilde{\chi}^{\pm}_{1,2}}$
                 & 3342, 6174 & 3504, 6303 \\

$m_{\tilde{g}}$  & 8295 & 14606 \\

\hline
$m_{\tilde{u}_{L,R}}$
                 & 8906, 11765 & 13188, 14206 \\

$m_{\tilde{t}_{1,2}}$
                 & 11621, 12954 & 14181, 14975 \\

\hline
$m_{\tilde{d}_{L,R}}$
                 & 11179, 11765 & 13450, 14205 \\

$m_{\tilde{b}_{1,2}}$
                 & 11264, 12955 & 14098, 14975 \\

\hline
$m_{\tilde{\nu}_{1,2}}$
                 & 6970, 7001 & 5180, 5306 \\

$m_{\tilde{\nu}_{3}}$
                 & 7002 & 5307 \\

\hline
$m_{\tilde{e}_{L,R}}$
                 & 4882, 6971 & 2264, 5181 \\

$m_{\tilde{\tau}_{1,2}}$
                 & 4968, 7002 & 2745, 5307 \\

\hline
\hline
\end{tabular}
}

\caption{Two representative benchmark points satisfying the GUT-scale
and gluino--squark hierarchy requirements. Point~1 belongs to the red
region and satisfies
$m_{\tilde g}<m_{\tilde u_L},m_{\tilde u_R},m_{\tilde t_1}$,
while Point~2 belongs to the yellow region and satisfies
$m_{\tilde g}>m_{\tilde u_L},m_{\tilde u_R},m_{\tilde t_1}$.
Sparticle and Higgs masses are given in GeV.
\label{table2}}
\end{table*}

For a stau NLSP, the corresponding chains terminate in
\begin{equation}
pp\to {\rm superpartners}\to
\widetilde\tau_1^+\widetilde\tau_1^-+X,
\qquad
\widetilde\tau_1\to\tau\widetilde G.
\label{eq:stau-cascade}
\end{equation}
Prompt decays give tau-rich final states with missing momentum, intermediate
lifetimes give displaced charged tracks or kinked trajectories, and a
sufficiently long-lived stau behaves as a heavy stable charged particle.  A
collider projection based only on $m_{\widetilde g}$ cannot cover all these
possibilities with one sensitivity curve.  This is the physical reason the $11$, and $17$~TeV lines are retained only as topology-dependent guides.

The red points $\rm M_{\rm GUT} > 0.84 \times 10^{16}$ GeV class is especially interesting because the $GUT$ condition at high scale and the NLSP identity become correlated.  In the present scan, red points are
predominantly prefer the neutralino-NLSP  and simultaneously occupy the lighter colored spectrum.  A search program combining jets plus missing
momentum with prompt and delayed photon categories is therefore very useful.

For the convenience of the readers, two representative benchmark points are listed in Table~\ref{table2}. The squark hierarchy determines which gluino reach should be used. For Point~1, $m_{\tilde g}=8295$~GeV while the up-type squark has $m_{\tilde u_L}=8906$~GeV;, $m_{\tilde t_{1}}=11264$ GeV and the $\rm M_{\rm GUT}=0.853 \times 10^{16}$ GeV. Thus point 1 represents red points where SUSY GUTS with GMSB  can be probed via gluino decays through heavy flavor at the future 100 TeV colliders such as $\rm LHC_{\rm hh}$ and SppC. On the other hand points with similar features but $\rm M_{\rm GUT}< 0.84 \times 10^{16}$ GeV can be probed at Hyper-Kamiokande experiment.    
 For Point~2, $m_{\tilde g}=14606$~GeV while the up-type squark has $m_{\tilde u_L}=13188$~GeV;, $m_{\tilde t_{1}}=14181$ GeV and the $\rm M_{\rm GUT}=0.592 \times 10^{16}$ GeV. Thus point 2 represents yellow points where SUSY GUTS with GMSB  can be probed via at Hyper-Kamiokande experiment and as well as gluino decays through light flavor at the future 100 TeV colliders such as $\rm LHC_{\rm hh}$ and SppC.

\section{Conclusion}

This paper is a continuation of our previous work in which we assumed the grand desert hypothesis and proposed to probe Supersymmetric Grand Unified Theories (GUTs) at the future proton-proton collider and Hyper-Kamiokande experiment. 
With the grand desert hypothesis, we proved that we can probe the Supersymmetric Grand Unified Theories (GUTs) with gauge meidation at the future proton-proton collider and Hyper-Kamiokande experiment. For the GUT scale $M_{\rm GUT}$ less than or equal to $0.84\times10^{16}$~GeV, the proton lifetime for dimension-six proton decay via heavy gauge boson exchange is less than or around $5\times10^{34}$ years, and thus we can probe it at the Hyper-Kamiokande experiment. For $ M_{\rm GUT} \ge 0.84 \times 10^{16}$ GeV, the lightest neutralino is the NLSP, as well as the upper mass bounds on the gluino and first two generation squarks are 8.64 TeV and 11.82 TeV, respectively. Therefore, the gluino and first two generation squarks can be probed at the future 100 TeV proton-proton collider such as ${\rm FCC}_{\rm hh}$ and SppC. 
Moreover, we discuss the mass spectra for the colored sparticles, sleptons, pseudo-scalar Higgs and electroweakinos. If the sparticles were discovered, it might give us hint on the messenger scale, the effective supersymmetry breaking scale, and the renormalization group evolution of the supersymmetry breaking soft terms.

\begin{acknowledgments}
SR wants to thanks TL and the Institute for Theoretical Physics, Chinese Academy of Sciences for the hospitality where most part of the work has been done.
TL is supported in part by the National Key Research and Development Program of China Grant No. 2020YFC2201504, by the Projects No. 11875062, No. 11947302, No. 12047503, and No. 12275333 supported by the National Natural Science Foundation of China, by the Key Research Program of the Chinese Academy of Sciences, Grant No. XDPB15, by the Scientific Instrument Developing Project of the Chinese Academy of Sciences, Grant No. YJKYYQ20190049, by the International Partnership Program of Chinese Academy of Sciences for Grand Challenges, Grant No. 112311KYSB20210012, and by the Henan Province Outstanding Foreign Scientist Studio Project, No.GZS2025008. This work was supported by the High Performance Computing Platform of Henan Normal University.

\end{acknowledgments}

\end{document}